\documentclass[preprint2]{aastex}

\slugcomment{Published in Astron. J. 141, 32}

\shorttitle{$BVRI$ photometry of 53 unusual asteroids}
\shortauthors{Ye}

\begin{document}

\title{$BVRI$ photometry of 53 unusual asteroids}

\author{Q.-z. Ye{1}}
\affil{Department of Atmospheric Sciences, School of Environmental Science and Engineering, Sun Yat-sen University, China}
              \email{tom6740@gmail.com}

\altaffiltext{1}{Present address: 404, 12 Huasheng St, Guangzhou, China}

\begin{abstract}
We present the results of $BVRI$ photometry and classification of 53 unusual asteroids, including 35 near-Earth asteroids (NEAs), 6 high eccentricity/inclination asteroids, and 12 recently-identified asteroid-pair candidates. Most of these asteroids were not reportedly classified prior to this work. For the few asteroids that have been previously studied, the results are generally in rough agreement. In addition to observe and classify these objects, we merge the results from several photometric/spectroscopic surveys to create a largest-ever sample with 449 spectrally classified NEAs for statistical analysis. We identify a ``transition point'' of the relative number of C/X-like and S-like NEAs at $H\sim18\Leftrightarrow D\sim1$km with confidence level at $\sim95$\% or higher. We find that the C/X-like:S-like ratio for $18\leq H<22$ is about two times higher than that of $H<18$ ($0.33\pm0.04$ versus $0.17\pm0.02$), virtually supporting the hypothesis that smaller NEAs generally have less weathered surface (therefore, less reddish appearance) caused by younger collision ages.
\end{abstract}


\section{Introduction}

The origin of unusual asteroid groups such as the near-Earth asteroids (NEAs) and paired-asteroids \citep{vok08,pra09} has raised many research interests over the recent years. A good way to understand this matter is to spectrally classify as many individuals as possible \citep[e.g.][]{bin96}. Since the 1980s, about a dozen of photometric and spectroscopic surveys aiming at determining taxonomic distributions of asteroids of different categories have been carried out (Table~\ref{tbl-1}) and derived spectral classifications for a few hundreds of NEAs as the result. However, among the $>7,500$ known NEAs, high eccentricity/inclination asteroids, and a few dozens proposed paired-asteroid candidates, the fraction of reportedly classified objects is still small (around 5\%), the fraction is even lower ($\sim1$\%) for those with measured physical characteristics (albedo, diameter, mineralogy, etc.).

Recent studies have suggested that the characteristics of sub-kilometer-size NEAs may be very different from that of kilometer-size NEAs \citep[see][for an overview]{tri10}, and a sign that has been noted for a decade is the tendency of overabundance of C/X-like (or neutral colored) asteroids among small size NEAs \citep{rab98}. Comparing with other issues, which generally require fine spectroscopic information and albedo measurement, the issue of overabundance of small C/X-like asteroids is relatively easy to work on, since only crude classification is required. However, confirmation is unable to be made by previous studies \citep[such as][]{dan03,bin04} due to the lack of data of sub-kilometer-size NEAs.

On the other hand, the recently identified paired-asteroid candidates are likely to be of common origin \citep{pra09}. It could be a convincing evidence to support this hypothesis if both components within a pair are proved to have identical classification, but however physical observations are lacking for almost all paired-asteroid candidates until now.

The method combining visual/near-infrared spectroscopy and thermal infrared measurement is preferred among all practical ground-based methods as it provides highest accuracy as well as most complete information of a target in most cases, but however it is also very time consuming and generally requires medium or large telescopes. By contrast, broad-band $BVRI$ photometry only allows crude classification, but it is more efficient than spectroscopy as it does not require as much times and efforts as the latter, and the result can be useful for preliminary diagnose purpose. In this study, we employed this method to investigate some selected unusual asteroids. Most of these asteroids had not been reportedly classified prior to the observational phase of this work. Description of the observation procedure, data reduction and details of classification is presented in Section 2 and 3. We then compared and merged our results with other reported studies to assess the result consistency between ours and others (Section 4.1) and investigate the degrees of consistency with theoretical expectations (Section 4.2 and 4.3).

\section{Observation}

The Lulin One-meter Telescope (LOT) at Lulin Observatory, Taiwan, was employed for this study, except for one asteroid, 2008 EV5, which was observed with the 0.41-m telescope of the same observatory. The 0.41-m telescope observations for 2008 EV5 were made in January 2009 with the $2048\times2048$ U42 CCD, while the LOT observations were all made during the observation runs in the dark period of January 2010 except (143651) 2003 QO104, which was observed in April 2009, with the $1340\times1300$ PI-1300B CCD and a $0.5\times$ focal reducer. A broad-band Bessell $BVRI$ filter system was used on both telescopes, with the wavelengths centered at 442, 540, 647 and 786 nm respectively.

Landolt standard stars \citep{lan92} are preferred in optical photometry because they can guarantee highest accuracy (better than 0.01-0.02 mag) in most cases. However, as Landolt standard stars are only available to a very limited region, in most occasions one needs to know the atmospheric extinction coefficient by observing standard stars in different airmass to work on targets locate far from Landolt fields. As we were unable to observe sufficient Landolt standard stars to get a secure extinction coefficient through every observing night, an alternative approach introduced by \citet{war07} is employed. Warner applied third-order polynomials on his optical observations of 128 carefully-chosen Landolt standard stars to find the conversion terms between the 2MASS $JHK$ system \citep{skr06} and the Landolt system, the errors of Warner's method are 0.034 for $B-V$ and $V-I$ and 0.021 for $V-R$ as indicated in his paper. The field-of-view for LOT and the 0.41-m telescope are $22'\times22'$ and $47'\times47'$ respectively, which are large enough to include sufficient 2MASS catalog stars for setting up a good in-field transformation. To assess the accuracy of Warner's method, we observed a few stars in M67 (NGC 2682) and derived their colors following the procedure described by Warner, then compared them with another high precision $VRI$ photometry \citep{tay08}. The resulting accuracy is better than 0.02 mag in $V-R$ and $V-I$ (Table~\ref{tbl-2}).

The targets are all observed with an airmass of $X\leq2$ with the predicted visual magnitude brighter than 19.0. Observational details as well as basic information of each target are shown in Tables~\ref{tbl-3} and~\ref{tbl-4} as for NEAs/high eccentricity (inclination) objects and paired-asteroid candidates. The exposure sequence for all targets is $B$-$R$-$V$-$I$ to minimize the error produced by significant brightness variation. Image frames are then bias-subtracted and flat-fielded, and the fringing effects in $I$-band images are also removed.

The raw observations are then inspected manually to exclude the bad cases, such as the target asteroid crossing over or passing very close to background stars, or low signal-to-ratio (SNR) caused by unstable weather conditions. Photometric measurements are then performed with Warner's software $MPO$ $Canopus$. At least ten background stars with known 2MASS magnitudes are used to derive the transformation coefficient between instrumental magnitude and standard magnitude for each field. In very few cases, the limited number of background stars cannot guarantee a good transform to be derived, so the coefficients derived from observations obtained in the same night with a similar airmass ($\Delta X<\sim0.05$) are used instead. Although each target was planned to be observed 3-5 times, various reasons (such as target/star encounter, unstable weather, and/or instrument problems) may prevent us to do so, and for some targets only one observation of each filter was obtained. These results shall be used with care.

\section{Classification}

Observations of 53 asteroids were obtained and reduced following the procedure described in Section 2, including 35 NEAs\footnote{We have also obtained $BV$ magnitude for (143947) 2003 YQ117 and $BVRI$ magnitude for (214088) 2004 JN13, suggesting R/S and A-type classification for these two objects, but however as the raw images suffered from bad seeing conditions resulting in uncertainty at the level of 0.2 mag, so these two measurements are excluded in our formal result.}, 6 high eccentricity/inclination asteroids, and 12 main-belt paired-asteroid candidates. The objects are then classified using the \citet{dan03}'s derivation of the Tholen taxonomy \citep{tho84}, intent for optical broad-band photometry\footnote{It should be noted that Dandy et al.'s derivation was used for their KPNO $B$ and Harris $VRI$ filter system, which is slightly different (the differences on wavelength centers are up to 2.5\%) from the Bessell $BVRI$ system applied in this study.}. We note that the spectral appearances of C-, B-, F-, G- and X-class are fairly close, making it difficult to classify them uniquely by broad-band photometry, so all objects with colors similar to these classes are considered as X-class. The only exception is that the object shown to be particularly blue ($B-V\ll0.75$) and can be tentatively considered as B-class with some confidence.

For each object, the second central moment about the $B-V$, $V-R$ and $V-I$ magnitudes from the typical colors for each taxonomic class are computed, and the class with minimum second central moment is assigned as the object's class. In some cases, the error range of the colors have covered more than one class and/or the second central moment of several classes are very close, so multiple classes are assigned to this object with the first class being most probable. In two cases [(13732) Woodall and (228747) 2002 VH3], only a very crude classification (C/X-like or S-like) can be made due to low-quality observations (see Section 4.3 for details). When the second central moment of the most probable class for a particular object is large ($\sim>0.003$), a ``$(u)$'' (uncertain) is appended to indicate that this classification may be uncertain.

The colors and the classifications for the observed NEAs/high eccentricity (inclination) asteroids and paired-asteroid candidates observed in this study are shown in Table~\ref{tbl-5} and Table~\ref{tbl-6} respectively. Classification and/or albedo measurement from previous work is also given if available. We note that a total of four objects were observed in different nights, color measurements across these nights with different calibration stars are found to be consistent. Measurements from each night are listed separately illustrate the accuracy and consistency of our work.

\section{Discussion}

\subsection{Comparisons with previous works on NEA colors}

Following the procedure described in Section 2 and 3, we have derived color indices and classifications for 35 NEAs, including 17 Amors, 13 Apollos and 5 Atens. Among the sample, a total of eight NEAs were reportedly classified by previous surveys; it has been found that our classifications are generally consistent with them. In addition, we note that a total of six NEAs in our sample were also observed by recently-conducted Warm-Spitzer program \citep{tri10}, with uncertainties around a factor of 2 unless otherwise specified (see Section 5.3.4 of their paper). Although Warm-Spitzer derives albedo of the NEAs and does not classify them directly, its observation can be a good addition to broad-band photometry, especially when the classification is ambiguous. Each case of these cross-observed NEAs is discussed below.

\paragraph{(5604) 1992 FE}

The $V-R$ and $V-I$ colors measured from our observation suggested V-, Q-, C- or R-type classification, with V-type to be most likely. This is consistent with the classification made by \citet{bin04} in spectroscopy. Albedo measurement by \citet{del03} and Warm-Spitzer observation \citep{tri10} also suggest a high albedo that support a V-type classification.

\paragraph{(5653) Camarillo}

The S- and Sq-type classification suggested by Dandy et al. and \citet{del10} are consistent with the S-type classification suggested by our observation.

\paragraph{(21088) 1992 BL2}

de Le\'{o}n et al. suggested a Sl-type classification which is consistent with the S-type classification suggested by our observation.

\paragraph{(38086) Beowulf}

Warm-Spitzer reported an albedo of 0.37 for this object. This is a rough match to our R-type classification according to the debiased mean albedo estimates for R-type NEAs given by \citet{stu04}, which is 0.340. However, we need to point out that Stuart \& Binzel was actually using the average albedo of main-belt asteroids derived from the work of \citet{ted02}, as no albedo measurements of any R-type NEAs had been reported (See Section 1.4, Paragraph 3 and Footnote 1 in their paper for details), hence this comparison can be misleading.

\paragraph{(54789) 2001 MZ7}

Near-infrared spectroscopy by \citet{laz05} reported (54789) 2001 MZ7 to be an X-complex asteroid while $BVRI$ photometry by \citet{bet10} suggested G-type classification. These two results are consistent with the X-type classification made in this study, since the degenerate X-type includes the C and G types. The color indices we measured also match with Betzler et al.'s within 0.03 mag.

\paragraph{(56048) 1998 XV39}

Our observation suggested an R-type classification while Sa-type suggested both by Lazzarin et al. and de Le\'{o}n et al. by spectroscopy. We note that (56048) 1998 XV39 appears redder than typical S-type asteroids in our observation ($0.770\pm0.045$ in $V-I$ magnitude versus the criteria of 0.889). As the redness of Sa-type asteroid is more close to that of R-type asteroid than to typical S-type asteroid \citep[see][Figure 15]{dem09}, it can be considered that our observation is in good consistency with the two spectroscopic observations.

\paragraph{(137805) 1999 YK5}

Our observation suggested an R- or Q-type classification which is inconsistent with the X-type classification by Binzel et al. by spectroscopy.

\paragraph{(137925) 2000 BJ19}

Our observation of this object suffered from low SNR, so the colors we measured cannot distinguish it from S-, T- or D-type classification, with S-type to be most likely. Binzel et al. suggested a Q-type classification for this object by spectroscopy, which is consistent with our S-type suggestion.

\paragraph{(138937) 2001 BK16}

Our observation showed that (138937) 2001 BK16 to be slightly blue, with $B-V$=$0.648\pm0.067$. Since the upper limit allowed by our error is $B-V$=0.71, which is within the range of X-class, we classify this object as X rather than the rare B-type. Warm-Spitzer observation yielded an albedo of 0.2, which can barely be matched with the albedo estimates for C- or X-type (debiased mean albedo of 0.101 for C-type and 0.072 for X-type as given by Stuart \& Binzel).

\paragraph{(143651) 2003 QO104}

Our observation suggested an R-type classification for (143651) 2003 QO104, this is consistent with the S-type suggested by Hicks and R- or S-type suggested by Betzler et al. \citep[see][]{bet10}. The color indices we measured matched with Betzler et al.'s within 0.02 mag. On the other hand, Warm-Spitzer observation indicated an albedo of 0.13 for this object. Considering the debiased mean albedo for R- and S-type NEAs given by Stuart \& Binzel to be 0.340 and 0.239 respectively, an S-type classification for this object might be more appropriate. However, as have mentioned earlier, we do not know the $true$ mean albedo for R-type NEAs, so the exact classification for this object is still an open question at this moment.

\paragraph{(159402) 1999 AP10}

Our observation suggested S-type classification for this object, but there is a $\sim0.1$ mag difference in $B-V$ magnitude from the typical color. Meanwhile, it has been reported that (159402) 1999 AP10 to be an L or S-type asteroid \citep{bet10} by spectroscopy and broad-band photometry. The difference between our and Betzler et al.'s observation of $B-V$ magnitude is about 0.12 mag, while $V-R$ and $V-I$ colors matched to 0.02 mag. Warm-Spitzer program gave an albedo estimate of 0.34, which does not oppose our S-type suggestion as the mean albedo for S-type NEAs was found to be 0.239 by Stuart \& Binzel. In addition, the error bar $\sigma$ for mean albedo determination of S-type NEAs is $\pm0.044$ as given by the authors, corresponding to an average dispersion of $\pm0.15$ considering there are 12 S-type NEAs used to determine the mean\footnote{As calculated by $D(x)=\sigma\sqrt{N-1}$ where $D(x)$ is the average dispersion.}, we can see that the Warm-Spitzer measurement mostly overlaps the albedo range of S-type NEAs.

\paragraph{(162998) 2001 SK162}

The $B-V$ and $V-R$ magnitudes derived from our observation suggested X- or V-type classification, while X-type is more likely. There is a rough consistency between our X-type suggestion and Binzel et al.'s T-type classification.

\paragraph{2002 LV}

Our observation suggests an X- or Q-type classification for this object, with Q-type to be more likely. On the other hand, observation from Warm-Spitzer suggested an albedo of 0.15. This does not resolve the ambiguity, since the uncertainty range covers the mean albedos for C-, X-, and Q-type NEAs by Stuart \& Binzel, which is 0.101, 0.072, and 0.247 respectively. However, considering that the average dispersions for each of the three complexes are 0.06, 0.06 and 0.15 respectively, we may conclude that a Q-type classification is the most probable classification for this object, since the uncertainties from both sources could have largest intersection under such a justification.

\paragraph{2007 MK13}

Colors derived from our observation fall between the typical colors of X- and T-type asteroids, with X-type classification to be more likely. Considering we have combined C- and X-class together, our classification is consistent with \citet{hic10}'s C-type classification.

\subsection{Statistical analysis with the results from other photometric and spectroscopic surveys}

To compare the similarities and differences between the results of some recently-conducted photometric and spectroscopic NEA surveys, we include the results from several surveys as listed in Table~\ref{tbl-1}, including \citet{del10}, SINEO \citep{laz05}, SMASS \citep{bus02,bin04}, \citet{dan03}, and \citet{ang02}\footnote{The result of \citet{fev07} is excluded as they are using a different taxonomy than Tholen's or Bus-DeMeo's.}. The classification results are firstly consolidated into several taxonomic complexes based on the scheme suggested by \citet{bin04} to allow the fractional abundances detected by each survey to be comparable (Table~\ref{tbl-7}). In addition, the taxonomic complexes are further consolidated into two general categories, ``C/X-like''\footnote{We see the need to use the ``C/X-like'' concept, instead of the ``C-like'' concept as used in many other papers to-date, to avoid interference over the ``dark/bright'' issue (which should primary rely on albedo information), although these two concepts are, in fact, equivalent with each other.} and ``S-like'', in order to determine the ratio of C/X-like and S-like observed by each survey. Considering the definition by \citet{mor02}\footnote{The classification method applied by Morbidelli et al. is actually dividing the asteroids into ``dark'' (corresponding to the ``C/X-like'' group in our study) and ``bright'' (corresponding to the ``S-like'' group in our study). However, some NEAs in their sample were classified based on taxonomic classification rather than real albedo. As the correlation between spectroscopic information and the target's albedo is still not well understood, we still apply a ``C/X-like versus S-like'' pattern that emphasizing the trend in color rather than a ``dark-bright'' pattern that emphasizes albedo. To simplify our following discussions, we also consider the results coming from a ``dark-bright'' pattern (as applied by the studies mainly based on spectroscopic data, such as Morbidelli et al.'s and Binzel et al.'s) to be quantitatively equivalent with the results from ``C/X-like versus S-like'' pattern.}, we consider the asteroids of class A, O, Q, R, S, U, and V as ``S-like'' while the asteroids of class C and D as ``C/X-like''. The degeneracy of X-complex is a problem since it includes members with diverse physical properties. As we don't have the fine physical data for each X-complex member, we consider the assumption of a dark-to-bright ratio (equivalent with our C/X-like:S-like ratio, as we have argued and presumed above) of 0.45 among X-complex NEAs as given by \citet{bin04} base on the albedo-taxonomy correlation of 22 X-complex NEAs. For the two photometric surveys (ours and Dandy et al.'s), things are more complicated since C- and X-complex cannot be distinguished, so we consider the relative number of C- and X-complex members among NEAs to be $\sim0.5$ as determined by Binzel et al., resulting a C/X-like:S-like ratio of $(0.5+0.45)/(1-0.45)\approx1.73$ in the combined C- and X-complex for the two photometric surveys. Finally, the objects with several possible classifications are excluded to avoid inducting further uncertainty.

As illustrated in Table~\ref{tbl-8}, the surveys agree on a dominate position of silicate composed asteroids (Q-, R-, S- and V-type). The fractions of each complex tend to be close on a larger sample $n$, suggesting that the fraction differences among the surveys are primary caused by random errors in observational sampling.

An interesting feature to look at is that the C/X-like:S-like ratio appeared to be magnitude dependent. The surveys with $\overline{H}<17$ (de Le\'{o}n et al.'s, Michelsen et al.'s, and Angeli \& Lazzaro's) all have the ratio smaller than 0.1, while the others all have the ratio larger than 0.1, suggesting a trend of more C/X-like asteroids at a larger $H$ (smaller size). This phenomenon has been noted by \citet{rab98}, \citet{mor02}, \citet{dan03} and \citet{bin04}, but no decisive conclusion were carried out due to lack of data among large $H$ (small size) asteroids. By contrast, Morbidelli et al. suggested that the ratio should slightly decrease with a larger $H$ due to observational bias effect and base on model prediction (see Section 2 of their paper), which is opposed to the implication of Table~\ref{tbl-8}.

To investigate this matter further, we merge all data together, creating a large dataset with 449 NEAs (with 434 NEAs with $H\leq22$ and 382 NEAs with $H\leq20$), and grouped them into one-magnitude-wide bins. For each magnitude bin, the C/X-like:S-like ratio is computed base on the scheme described above. The result is listed in Table~\ref{tbl-9}.

Comparing the result with Morbidelli et al.'s study, which has included 183 NEAs with $H<20$, the ratio variation from magnitude to magnitude between the two studies is similar: a lower C/X-like:S-like ratio on $H<18$ and a higher ratio on $H\geq18\Leftrightarrow D\leq1$km, which agrees the implication of Table ~\ref{tbl-8}. To check the statistical significance of this phenomenon, we divide the whole sample into two groups by an $H=18$ cutoff, with the sample $n$ of each group to be 246 ($H<18$) and 203 ($H\geq18$) respectively, and perform a $\chi^{2}$ test on the two groups. It has been shown that the distributions of two groups are different at a confidence level of 99.5\%, which is large enough to be considered as statistically significant. Although our treatment of the C/X-like:S-like ratio of the X-complex may induct some uncertainty, we note that even we consider a $1\sigma$ uncertainty of the numbers of X-complex members to be ``C/X-like'' (assuming random observation errors) for both $H<18$ and $H\geq18$, for which the numbers of X-complex members considered as ``C/X-like'' would be $10\pm3$ and $17\pm4$ respectively, the minimum possible confidence level is still shown to be $\sim95$\%. What is more, we notice that the C/X-like:S-like ratio we derived is $0.17\pm0.02$ for NEAs with $H<18$, which is in a good match of the model prediction of 0.18 according to Morbidelli et al, implying our treatment of the X-complex members with $H<18$ would lead to a good match with the model. However, things are no longer that case when it comes to an $H<20$ cutoff line, for which our estimate of the C/X-like:S-like ratio is shown to be $0.22\pm0.02$, about 3-7\% higher than Morbidelli et al.'s model prediction (0.17). It is noteworthy that we have to consider $all$ X-complex members with $18\leq H<20$ to be ``S-like'' as in an extreme situation for a compatible ratio with the model prediction (0.18 versus the model's $0.165\pm0.015$). In addition, our estimate of C/X-like:S-like ratio for $18\leq H<22$ (sample $n=187$ comparing with $n\sim50$ for $18\leq H<20$ in Morbidelli et al.'s sample) is $0.33\pm0.04$. In all, it is plausible to say that a transition point of C/X-like:S-like ratio exists at $H=18$ in our sample, which is consistent with the hypothesis that small asteroids generally have younger collision age and tend to be less weathered (thus, less reddish/neutral colored) than large ones \citep{bin98}, hence leads to the overabundance of C/X-like asteroids in small NEAs.

Morbidelli et al. suggested a factor of $\sim1.4$ between observed and true C/X-like:S-like ratio caused by the phase angle effect at discovery. Assuming that this number remains quasi-constant for different $H$, we will get a true C/X-like:S-like ratio of 0.24 for asteroids with $H<18$, which is in a good match with Morbidelli et al.'s model ($0.25\pm0.02$). However, for $H\geq18$, the true ratio can be as high as 0.5 in our sample.

\subsection{Discussion of the paired-asteroid candidate observations}

Because of the constraints of the target observability, only two pairs were with both components observed, they are (1979) Sakharov vs. (13732) Woodall and (11842) Kap'bos vs. (228747) 2002 VH3 respectively. Unfortunately, we were unable to obtain complete color indices for either pairs: for the first pair, (13732) Woodall is suffered from instrument problems so the $V-I$ magnitude is missing, while (228747) 2002 VH3 was just passing near a 4.5 mag star when the observation was taken for the second pair, so we are only able to make a very crude classification for these two objects. In general, both components of both pairs are classified as S-like asteroid, providing a weak evidence of the common-origin hypothesis as expected.

\section{Conclusion}

Our efforts have added the spectral data of some tens of interesting asteroids into knowledge. For the few asteroids that had been reportedly classified, our results are generally in rough agreement, indicating the goodness of our work.

Further to observing and classifying the asteroids, we also examined the matter of overabundance of C/X-like (neutral colored) asteroids among small size NEAs, a phenomenon that had previously been noted by several studies without a decisive conclusion. In a largest-ever sample we created from merging results from several photometric/spectroscopic surveys with 449 spectrally classified NEAs in total, we identify a ``transition point'' of the C/X-like:S-like ratio among NEAs at $H\sim18\Leftrightarrow D\sim1$km with statistical confidence level at 95\% or higher. The C/X-like:S-like ratio for NEAs with $H<18$ is estimated to be $0.17\pm0.02$ as measured in this sample, which is in good match with previous model prediction by Morbidelli et al.; while the ratio for NEAs with $18\leq H<22$ is estimated to be $0.33\pm0.04$, about two times higher than that for $H<18$ and being inconsistent with model prediction. However, this finding supports the hypothesis that asteroids with less reddish appearances are more abundant among small size NEAs than large ones due to younger collision ages and less weathered surface.

\acknowledgments

The author is very grateful to Alan Harris for his valuable help and constructive comments which have led to an advancement of this work. The author is also very grateful to Alberto Betzler, Wing-Huen Ip, Timophy Spahr and Jin Zhu for their help and encouragements to author to initiate this work; and to Johnson Lau, Petr Pravec, Brian Warner and Linjiong Zhou for their help on data processing and manuscript preparation. In addition, the author thanks Hung-Chin Lin, Chi Sheng Lin and Hsiang-Yao Hsiao from Lulin Observatory on behalf of the Graduate Institute of Astronomy, National Central University, for providing assistance on obtaining the observational data and granting the permission to use their licensed copy of \textit{MPO Canopus} for data processing. This work is supported by a 2007 Gene Shoemaker NEO Grant from the Planetary Society. The LOT and 0.41-m telescope are operated with support from the Graduate Institute of Astronomy, National Central University.

\clearpage







\clearpage

\begin{deluxetable}{c c c c}
\tabletypesize{\scriptsize}
\rotate
\tablecaption{Some photometric/spectroscopic NEA surveys conducted over the recent decade\label{tbl-1}}
\tablewidth{0pt}
\tablehead{
\colhead{Publication} & \colhead{Observation period} & \colhead{NEA sample size} & \colhead{Method}
}
\startdata
            \citet{del10} & 2002-2007 & 77 & VIS/NIR\tablenotemark{a} spectroscopy \\
            \citet{fev07} & 1997-1998, 2004 & 55 & VIS\tablenotemark{b} spectroscopy \\
            \citet{mic06} & 2003 & 12 & VIS spectroscopy \\
            SINEO \citep{laz05} & 2002-2004 & 36 & VIS/NIR spectroscopy \\
            SMASS \citep{bin04,bus02} & 1994-2002 & 310 & VIS spectroscopy \\
            \citet{dan03} & 2000-2001 & 56 & Broad-band $BVRIZ$ photometry \\
            $S^{3}OS^{2}$ \citep{ang02} & 1996-2001 & 12 & VIS spectroscopy \\
\enddata
         \tablenotetext{a}{VIS stands for visual channel (about 390-750nm) while NIR stands for near-infrared channel (about 700-3,000nm)}
         \tablenotetext{b}{Although the study is labeled as visual and near-infrared spectroscopy in the publication, it only covers the near-infrared channel to ~1,000nm, so it is considered as visual spectroscopy in this summary.}
\end{deluxetable}

\begin{deluxetable}{c c c c c c c}
\tabletypesize{\scriptsize}
\rotate
\tablecaption{Comparison on the $VRI$ photometry of M67 (NGC 2682) between this work and \citet{tay08}\label{tbl-2}}
\tablewidth{0pt}
\tablehead{
\colhead{Star} & \colhead{$V-R$ by Taylor et al.} & \colhead{$V-I$ by Taylor et al.} & \colhead{Date observed} & \colhead{Airmass} & \colhead{$V-R$ by this study} & \colhead{$V-I$ by this study}
}
\startdata
            NGC 2682 F132 & $0.350\pm0.003$ & $0.679\pm0.006$ & 2010-01-12 & 1.69 & $0.336\pm0.011$ & $0.676\pm0.023$ \\
             &  &  & 2010-01-13 & 1.34 & $0.342\pm0.003$ & $0.677\pm0.007$ \\
            NGC 2682 I51 & $0.336\pm0.003$ & $0.671\pm0.007$ & 2010-01-12 & 1.69 & $0.331\pm0.016$ & $0.651\pm0.024$ \\
             &  &  & 2010-01-13 & 1.34 & $0.334\pm0.007$ & $0.655\pm0.010$ \\
\enddata
\end{deluxetable}

\begin{deluxetable}{c c c c c c c c}
\tablewidth{0pt}
\tablecaption{Observational details of the observed NEAs and high eccentricity/inclination objects\label{tbl-3}}
\tablehead{
\colhead{Object} & \colhead{$H$} & \colhead{Orbit} & \colhead{Date observed}  &
\colhead{$V_{obs}$} & \colhead{Exp. time} & \colhead{Cycle} & \colhead{Tot. time}
}
\startdata
            (5604) 1992 FE & 16.4 & ATE & 2010-1-13 & 19.4 & 60s & 5 & 40min \\
            (5653) Camarillo & 15.4 & AMO & 2010-1-13 & 18.5 & 60s & 5 & 40min \\
            (16868) 1998 AK8 & 16.6 & UNU & 2010-1-13 & 17.3 & 30s & 3 & 16min \\
            (20898) Fountainhills & 11.0 & UNU & 2010-1-8 & 14.7 & 60s & 5 & 39min \\
            (21088) 1992 BL2 & 14.2 & AMO & 2010-1-8,13 & 17.4,17.4 & 60s,60s & 5,5 & 39min,40min \\
            (24761) Ahau & 17.4 & APO & 2010-1-8 & 15.6 & 20s & 3 & 15min \\	
            (35432) 1998 BG9 & 19.3 & AMO & 2010-1-13 & 18.9 & 30s & 3 & 24min \\
            (38086) Beowulf & 17.1 & APO & 2010-1-13 & 18.2 & 30s & 2 & 11min \\
            (54789) 2001 MZ7 & 14.8 & AMO & 2010-1-8 & 14.6 & 30s & 5 & 30min \\
            (66251) 1999 GJ2 & 17.0 & AMO & 2010-1-13 & 18.2 & 90s & 2 & 18min \\
            (68216) 2001 CV26 & 16.3 & APO & 2010-1-8 & 17.6 & 30s & 3 & 22min \\
            (86067) 1999 RM28 & 16.4 & AMO & 2010-1-8 & 17.1 & 30s & 3 & 17min \\
            (96177) 1984 BC & 16.0 & UNU & 2010-1-13 & 17.7 & 2min & 5 & 58min \\
            (99248) 2001 KY66 & 16.2 & APO & 2010-1-13 & 19.0 & 90s & 4 & 39min \\
            (103067) 1999 XA143 & 16.6 & APO & 2010-1-8 & 16.8 & 30s & 5 & 30min \\
            (122180) 2001 KV43 & 17.3 & AMO & 2010-1-10,12 & 18.1,18.3 & 3min,3min & 5,2 & 56min,37min \\
            (137805) 1999 YK5 & 16.6 & ATE & 2010-1-8 & 17.5 & 30s & 5 & 30min \\
            (137925) 2000 BJ19 & 15.8 & APO & 2010-1-10,12 & 19.0,18.7 & 60s & 5,5 & 22min,14min \\
            (138937) 2001 BK16 & 17.3 & APO & 2010-1-10,12 & 17.4,17.7 & 30s,30s & 5,3 & 22min,7min \\
            (143651) 2003 QO104 & 16.0 & AMO & 2009-4-24 & 16.6 & 60s & 9 & 59min \\
            (152742) 1998 XE12 & 19.1 & ATE & 2010-1-10,12 & 18.2,18.0 & 30s,30s & 5,5 & 22min,14min \\
            (159402) 1999 AP10 & 16.0 & AMO & 2010-1-8 & 15.7 & 30s & 5 & 30min \\
            (162566) 2000 RJ34 & 15.2 & AMO & 2010-1-12 & 18.9 & 60s & 5 & 39min \\
            (162998) 2001 SK162 & 17.9 & AMO & 2010-1-10 & 18.6 & 60s & 3 & 17min \\
            (230111) 2001 BE10 & 19.1 & ATE & 2010-1-10,12 & 17.7,17.7 & 20s,20s & 1,3 & 2min,28min \\
            2002 LV & 16.6 & APO & 2010-1-9 & 18.1 & 60s & 5 & 40min \\
            2003 SM4 & 15.4 & UNU & 2010-1-9 & 19.0 & 90s & 3 & 28min \\
            2004 NZ8 & 16.1 & UNU & 2010-1-9 & 18.3 & 2min & 5 & 58min \\
            2004 TB18 & 17.8 & APO & 2010-1-12 & 18.2 & 20s & 4 & 21min \\
            2004 XD50 & 18.5 & AMO & 2010-1-9 & 17.6 & 30s & 4 & 24min \\
            2005 EN36 & 17.2 & UNU & 2010-1-8,9 & 17.5,17.5 & 2min,2min & 1,5 & 6min,58min \\
            2005 MC & 16.6 & AMO & 2010-1-8 & 16.5 & 30s & 3 & 18min \\
            2006 UR & 19.4 & AMO & 2010-1-8 & 16.4 & 30s & 2 & 11min \\
            2006 YT13 & 18.4 & APO & 2010-1-10 & 18.1 & 20s & 5 & 20min \\		
            2007 MK13 & 20.0 & APO & 2010-1-8 & 17.2 & 30s & 3 & 43min \\
            2007 UR3 & 21.2 & AMO & 2010-1-9 & 17.7 & 60s & 5 & 39min \\
            2008 EV5 & 20.0 & ATE & 2009-1-4 & 14.7 & 20s & 5 & 10min \\
            2008 YZ32 & 20.1 & APO & 2010-1-9 & 17.3 & 20s & 5 & 15min \\
            2009 UV18 & 16.0 & AMO & 2010-1-9 & 18.3 & 30s & 5 & 30min \\
            2009 XR2 & 18.6 & AMO & 2010-1-9 & 16.5 & 30s & 3 & 17min \\
            2010 AE30 & 23.6 & APO & 2010-1-9 & 18.5 & 60s & 2 & 39min \\
\enddata
         \tablecomments{The absolute magnitude $H$ for each object is obtained from the Minor Planet Center Orbit Database (MPCORB). The orbit types are classified as follows: Main-belt Asteroids (MBA), Atens (ATE), Apollos (APO), and Amors (AMO). Asteroids with $e>0.4$ and/or $i>40^{\circ}$ but not fall into the category of NEAs ($q<1.3$) are classified as unusual objects (UNU). The visual magnitude $V_{obs}$ given here is the mean value of the $V$-band observations.\\}
\end{deluxetable}

\begin{deluxetable}{c c c c c c c c}
\tabletypesize{\scriptsize}
\rotate
\tablecaption{Observational details of the observed paired-asteroid candidates\label{tbl-4}}
\tablewidth{0pt}
\tablehead{
\colhead{Object} & \colhead{$H$} & \colhead{Companion} & \colhead{Date observed}  &
\colhead{$V_{obs}$} & \colhead{Exp. time} & \colhead{Cycle} & \colhead{Tot. time}
}
\startdata
            (1979) Sakharov & 13.5 & (13732) Woodall\tablenotemark{a} & 2010-1-12,13 & 18.8,18.8 & 90s,90s & 1,1 & 7min,5min \\
            (2110) Moore-Sitterly & 13.8 & (44612) 1999 RP27 & 2010-1-10 & 18.4 & 2min & 5 & 42min \\		
            (4765) Wasserburg & 14.1 & 2001 XO105 & 2010-1-9 & 16.9 & 60s & 3 & 23min \\
            (5026) Martes & 12.9 & 2005 WW113 & 2010-1-10 & 18.6 & 60s & 1 & 11min \\		 
            (11842) Kap'bos & 13.8 & (228747) 2002 VH3\tablenotemark{a} & 2010-1-9 & 18.5 & 60s & 5 & 39min \\
            (13732) Woodall & 14.4 & (1979) Sakharov\tablenotemark{a} & 2010-1-10 & 18.7 & 90s & 4 & 29min \\		
            (15107) Toepperwein & 14.3 & 2006 AL54 & 2010-1-10,12 & 18.5,18.5 & 2min,2min & 5,2 & 43min,9min \\
            (25884) 2000 SQ4 & 14.6 & (48527) 1993 LC1 & 2010-1-10 & 18.5 & 150s & 2 & 14min \\		
            (52852) 1998 RB75 & 14.6 & 2003 SC7 & 2010-1-12 & 18.7 & 5min & 5 & 115min \\
            (54041) 2000 GQ113 & 14.5 & 2002 TO134 & 2010-1-10 & 18.6 & 2min & 2 & 12min \\
            (56048) 1998 XV39 & 15.0 & (76148) 2000 EP17 & 2010-1-10,12 & 17.4,18.4 & 3min,3min & 5,5 & 54min,37min \\
            (228747) 2002 VN3 & 16.5 & (11842) Kap'bos\tablenotemark{a} & 2010-1-9 & 18.8 & 2min & 3 & 33min \\
\enddata
\tablecomments{The absolute magnitude $H$ for each object is obtained from the Minor Planet Center Orbit Database (MPCORB). The visual magnitude $V_{obs}$ given here is the mean value of the $V$-band observations.\\}
         \tablenotetext{a}{The companion of this object is also observed in this study.}
\end{deluxetable}

\begin{deluxetable}{c c c c c c c c c}
\tabletypesize{\scriptsize}
\rotate
\tablecaption{Photometry and classification results of observed NEAs and high eccentricity/inclination objects\label{tbl-5}}
\tablewidth{0pt}
\tablehead{
\colhead{Object} & \colhead{Date observed} & \colhead{$V_{obs}$} & \colhead{Class (this work)} & \colhead{Previous class}  &
\colhead{$B-V$} & \colhead{$V-R$} & \colhead{$V-I$} & \colhead{Albedo}
}
\startdata
            (5604) 1992 FE & 2010-1-13 & 19.4 & VQXR & V\tablenotemark{a} & & $0.439\pm0.077$ & $0.679\pm0.083$ & 0.38-0.69\tablenotemark{b} \\
            (5653) Camarillo & 2010-1-13 & 18.5 & S & Sq\tablenotemark{c} & $0.985\pm0.062$ & $0.465\pm0.050$ & $0.893\pm0.048$ & \\
            (16868) 1998 AK8 & 2010-1-13 & 17.3 & S & & $0.908\pm0.031$ & $0.443\pm0.014$ & $0.857\pm0.002$ & \\
            (20898) Fountainhills & 2010-1-8 & 14.7 & X & & $0.767\pm0.008$ & $0.428\pm0.010$ & $0.826\pm0.008$ & \\
            (21088) 1992 BL2 & 2010-1-8 & 17.4 & S & Sl\tablenotemark{c} & $0.972\pm0.048$ & $0.490\pm0.006$ & $0.913\pm0.015$ & \\
             & 2010-1-13 & 17.4 & & & $0.919\pm0.055$ & $0.438\pm0.014$ & $0.907\pm0.028$ & \\
            (24761) Ahau & 2010-1-8 & 15.6 & S & & $0.835\pm0.023$ & $0.469\pm0.008$ & $0.872\pm0.002$ & \\
            (35432) 1998 BG9 & 2010-1-13 & 18.9 & A & & & $0.547\pm0.030$ & $0.913\pm0.045$ & \\
            (38086) Beowulf & 2010-1-13 & 18.2 & R$(u)$ & & $0.999\pm0.020$ & $0.447\pm0.018$ & $0.645\pm0.098$ & 0.37\tablenotemark{b} \\
            (54789) 2001 MZ7 & 2010-1-8 & 14.6 & X & X\tablenotemark{d}, G\tablenotemark{e} & $0.707\pm0.011$ & $0.389\pm0.004$ & $0.711\pm0.002$ & \\
            (66251) 1999 GJ2 & 2010-1-13 & 18.2 & S & Sa\tablenotemark{c}, Sa\tablenotemark{e} & $0.976\pm0.088$ & $0.448\pm0.036$ & $0.844\pm0.021$ & \\
            (68216) 2001 CV26 & 2010-1-8 & 17.6 & R & & $0.916\pm0.036$ & $0.479\pm0.007$ & $0.811\pm0.025$ & \\
            (86067) 1999 RM28 & 2010-1-8 & 17.1 & S$(u)$ & & $0.996\pm0.069$ & $0.499\pm0.018$ & $0.881\pm0.021$ & \\
            (96177) 1984 BC & 2010-1-13 & 17.7 & S & & $0.864\pm0.045$ & $0.484\pm0.018$ & $0.921\pm0.019$ & \\
            (99248) 2001 KY66 & 2010-1-13 & 19.0 & ST & & $0.865\pm0.048$ & $0.455\pm0.041$ & $0.998\pm0.072$ & \\
            (103067) 1999 XA143 & 2010-1-8 & 16.8 & S & & $0.925\pm0.029$ & $0.458\pm0.011$ & $0.878\pm0.009$ & \\
            (122180) 2001 KV43 & 2010-1-10 & 18.1 & S & & $0.894\pm0.067$ & $0.508\pm0.036$ & & \\
             & 2010-1-12 & 18.3 & & & & & $0.915\pm0.072$ & \\
            (137805) 1999 YK5 & 2010-1-8 & 17.5 & RQ & X\tablenotemark{a} & $0.908\pm0.035$ & $0.390\pm0.051$ & $0.704\pm0.036$ & \\
            (137925) 2000 BJ19 & 2010-1-10 & 19.0 & STD & Q\tablenotemark{a} & $0.756\pm0.135$ & $0.508\pm0.079$ & & \\
             & 2010-1-12 & 18.7 & & & & & $0.915\pm0.072$ & \\
            (138937) 2001 BK16 & 2010-1-10 & 17.4 & X & & $0.648\pm0.067$ & $0.369\pm0.034$ & & 0.21\tablenotemark{b} \\
             & 2010-1-12 & 17.7 & & & & & $0.565\pm0.082$ & \\
            (143651) 2003 QO104 & 2009-4-24 & 16.6 & R & RS\tablenotemark{e} & $0.903\pm0.008$ & $0.454\pm0.011$ & $0.797\pm0.019$ & \\
            (152742) 1998 XE12 & 2010-1-10 & 18.2 & RS & & $0.898\pm0.100$ & $0.460\pm0.062$ & & \\
             & 2010-1-12 & 18.0 & & & & & $0.825\pm0.043$ & \\
            (159402) 1999 AP10 & 2010-1-8 & 15.7 & S$(u)$ & S\tablenotemark{f} & $0.983\pm0.023$ & $0.485\pm0.007$ & $0.868\pm0.009$ & 0.34\tablenotemark{b} \\
            (162566) 2000 RJ34 & 2010-1-12 & 18.9 & X & & $0.755\pm0.136$ & $0.400\pm0.032$ & $0.836\pm0.047$ & \\
            (162998) 2001 SK162 & 2010-1-10 & 18.6 & XV & T\tablenotemark{a} & $0.822\pm0.139$ & $0.386\pm0.018$ & & \\
            (230111) 2001 BE10 & 2010-1-10 & 17.7 & R & & $0.929$ & $0.420$ & & \\
             & 2010-1-12 & 17.7 & & & $0.982\pm0.087$ & $0.496\pm0.035$ & $0.870\pm0.095$ & \\
            2002 LV & 2010-1-9 & 18.1 & QX & & $0.772\pm0.144$ & $0.425\pm0.039$ & $0.719\pm0.046$ & 0.15\tablenotemark{b} \\
            2003 SM4 & 2010-1-9 & 19.0 & V & & & $0.387\pm0.052$ & $0.602\pm0.055$ & \\
            2004 NZ8 & 2010-1-9 & 18.3 & X & & $0.704\pm0.085$ & $0.404\pm0.027$ & $0.707\pm0.028$ & \\
            2004 TB18 & 2010-1-12 & 18.2 & AS & & & $0.510\pm0.084$ & $1.000\pm0.138$ & \\
            2004 XD50 & 2010-1-9 & 17.6 & R & & $0.927\pm0.030$ & $0.458\pm0.005$ & $0.791\pm0.025$ & \\
            2005 EN36 & 2010-1-8 & 17.5 & X & & $0.701$ & $0.356$ & $0.697$ & \\
             & 2010-1-9 & 17.5 & & & $0.737\pm0.016$ & $0.369\pm0.010$ & $0.681\pm0.014$ & \\
            2005 MC & 2010-1-8 & 16.5 & X & & $0.752\pm0.022$ & $0.413\pm0.010$ & $0.766\pm0.010$ & \\
            2006 UR & 2010-1-8 & 16.4 & R & & $0.969\pm0.097$ & $0.434\pm0.060$ & $0.754\pm0.053$ & \\
            2006 YT13 & 2010-1-10 & 18.1 & AR & & $1.042\pm0.088$ & $0.475\pm0.057$ & & \\		
            2007 MK13 & 2010-1-8 & 17.2 & XT & C\tablenotemark{f} & $0.783\pm0.016$ & $0.414\pm0.010$ & $0.838\pm0.019$ & \\
            2007 UR3 & 2010-1-9 & 17.7 & X & & $0.631\pm0.054$ & $0.381\pm0.025$ & $0.603\pm0.044$ & \\
            2008 EV5 & 2009-1-4 & 14.7 & X & & $0.722\pm0.033$ & $0.363\pm0.010$ & $0.702\pm0.010$ & \\
            2008 YZ32 & 2010-1-9 & 17.3 & B & & $0.535\pm0.032$ & $0.207\pm0.055$ & $0.507\pm0.012$ & \\
            2009 UV18 & 2010-1-9 & 18.3 & A & & $0.967\pm0.094$ & $0.520\pm0.019$ & $0.957\pm0.052$ & \\
            2009 XR2 & 2010-1-9 & 16.5 & R & & $0.908\pm0.017$ & $0.484\pm0.008$ & $0.724\pm0.031$ & \\
            2010 AE30 & 2010-1-9 & 18.5 & S & & $0.862\pm0.090$ & $0.461\pm0.010$ & $0.845\pm0.011$ & \\
\enddata
         \tablenotetext{a}{\citet{bus02,bin04}}
         \tablenotetext{b}{\citet{tri10}}
         \tablenotetext{c}{\citet{del10}}
         \tablenotetext{d}{\citet{laz05}}
         \tablenotetext{e}{\citet{bet10}}
         \tablenotetext{f}{\citet{hic10}}
         \tablecomments{The visual magnitude $V_{obs}$ given here is the mean value of the $V$-band observations. Albedos given by reference b all have uncertainties around a factor of 2.\\}
\end{deluxetable}

\begin{deluxetable}{c c c c c c c c}
\tabletypesize{\scriptsize}
\rotate
\tablecaption{Photometry and classification results of observed paired-asteroid candidates\label{tbl-6}}
\tablewidth{0pt}
\tablehead{
\colhead{Object} & \colhead{Companion} & \colhead{Date observed} & \colhead{$V_{obs}$} & \colhead{Class (this work)} & \colhead{$B-V$} &
\colhead{$V-R$} & \colhead{$V-I$}
}
\startdata
            \textbf{(1979) Sakharov} & \textbf{(13732) Woodall} & \textbf{2010-1-12} & \textbf{18.8} & \textbf{S$(u)$} & & \textbf{$0.407$} & \textbf{$0.922$} \\
             & & \textbf{2010-1-13} & \textbf{18.8} & & \textbf{$1.086$} & \textbf{$0.411$} & \\
            \textbf{(13732) Woodall} & \textbf{(1979) Sakharov} & \textbf{2010-1-10} & \textbf{18.7} & \textbf{S-like} & \textbf{$0.864\pm0.147$} & \textbf{$0.468\pm0.068$} & \\	
            (2110) Moore-Sitterly & (44612) 1999 RP27 & 2010-1-10 & 18.4 & R & $0.956\pm0.034$ & $0.517\pm0.033$ & \\		
            (4765) Wasserburg & 2001 XO105 & 2010-1-9 & 16.9 & R & $0.852\pm0.043$ & $0.456\pm0.023$ & $0.813\pm0.040$ \\
            (5026) Martes & 2005 WW113 & 2010-1-10 & 18.6 & SRQV & $0.863\pm0.042$ & $0.440\pm0.047$ & \\		
            \textbf{(11842) Kap'bos} & \textbf{(228747) 2002 VH3} & \textbf{2010-1-9} & \textbf{18.5} & \textbf{R} & \textbf{$1.011\pm0.088$} & \textbf{$0.453\pm0.013$} & \textbf{$0.792\pm0.027$} \\
            \textbf{(228747) 2002 VH3} & \textbf{(11842) Kap'bos} & \textbf{2010-1-9} & \textbf{18.8} & \textbf{S-like} & \textbf{$0.704\pm0.154$} & \textbf{$0.480\pm0.057$} & \textbf{$0.829\pm0.111$}
             \\
            (15107) Toepperwein & 2006 AL54 & 2010-1-10 & 18.5 & A$(u)$ & $0.876\pm0.079$ & $0.492\pm0.027$ & \\
             & & 2010-1-12 & 18.5 & & & & $1.016\pm0.021$ \\
            (25884) 2000 SQ4 & (48527) 1993 LC1 & 2010-1-10 & 18.5 & XT & $0.709\pm0.055$ & $0.432\pm0.014$ & \\		
            (52852) 1998 RB75 & 2003 SC7 & 2010-1-12 & 18.7 & S & $0.891\pm0.083$ & $0.488\pm0.035$ & $0.908\pm0.079$ \\
            (54041) 2000 GQ113 & 2002 TO134 & 2010-1-10 & 18.6 & SRQ & $0.871\pm0.073$ & $0.459\pm0.027$ & \\
            (56048) 1998 XV39 & (76148) 2000 EP17 & 2010-1-10 & 17.4 & R & $0.874\pm0.039$ & $0.476\pm0.035$ & \\
             & & 2010-1-12 & 18.4 & & & & $0.770\pm0.045$ \\
\enddata
         \tablecomments{The visual magnitude $V_{obs}$ given here is the mean value of the $V$-band observations. The pairs those with both companions observed are listed in bold characters.\\}
\end{deluxetable}

\begin{table}
      \caption{Summary of consolidated taxonomic classes bases on the works of \citet{bin04} and \citet{stu04}}
         \label{tbl-7}
         \centering
         \begin{tabular}{c c}
            \hline\hline
            Taxonomic complex & Includes \\
            \hline	
            A & A \\
            C & C, Cb, Cg, Cgh, Ch, B, F, G \\
            D & D,T \\
            O & O \\
            Q & Q, Sq \\
            R & R \\
            S & S, Sa, Sk, Sl, Sr, Sv, K, L, Ld \\
            U & U\\
            V & V\\
            X & X, Xc, Xe, Xk, E, M, P \\
            \noalign{\smallskip}
            \hline
         \end{tabular}
\end{table}

\begin{deluxetable}{c c c c c c c c}
\tabletypesize{\scriptsize}
\rotate
\tablecaption{Fractional abundances of each taxonomic complex and apparent C/X-like:S-like ratio as analyzed from the results of several photometric/spectroscopic studies\label{tbl-8}}
\tablewidth{0pt}
\tablehead{
\colhead{Taxonomic complex} & \colhead{This work} & \colhead{\citet{del10}\tablenotemark{a}} & \colhead{\citet{mic06}} & \colhead{\citet{laz05}} &
\colhead{\citet{bin04}\tablenotemark{a}} & \colhead{\citet{dan03}} & \colhead{\citet{ang02}}
}
\startdata
            Method & Photometry & Spectroscopy & Spectroscopy & Spectroscopy & Spectroscopy & Photometry & Spectroscopy \\
            Sample $n$ & 25 & 77 & 12 & 36 & 310 & 51 & 12 \\
            $\overline{H}$ & 17.2 & 16.1 & 15.7 & 17.6 & 17.8 & 17.4 & 14.6 \\
            A & 8\% & 3\% & 0\% & 3\% & $\sim0$\% & 0\%	8\% \\
            C & - & 1\% & 0\% & 11\% & 7\% & 34\%\tablenotemark{b} & 0\% \\
            D & 0\% & 0\% & 8\% & 0\% & 3\% & 0\% & 0\% \\
            O & 0\% & 4\% & 0\% & 0\% & 2\% & 0\% & 0\% \\
            Q & 0\% & 25\% & 0\% & 8\% & 26\% & 25\% & 17\% \\
            R & 28\% & 0\% & 0\% & 0\% & $\sim0$\% & 14\% & 0\% \\
            S & 36\% & 58\% & 92\% & 58\% & 40\% & 22\% & 58\% \\
            U & 0\% & 0\% & 0\% & 0\% & 1\% & 0\% & 0\% \\
            V & 0\% & 8\% & 0\% & 0\% & 5\% & 2\% & 8\% \\
            X & 24\%\tablenotemark{b} & 3\% & 0\% & 11\% & 15\% & - & 8\% \\
            C/X-like:S-like ratio & 0.14 & 0.01 & 0.09 & 0.27 & 0.21 & 0.21 & 0.04\\
\enddata
         \tablenotetext{a}{Mars-crossers in the sample have been removed in this comparison.}
         \tablenotetext{b}{The fraction of C-complex includes the contribution from B-, C-, E- and X-complex.}
\end{deluxetable}

\begin{table}
      \caption{Counts of C/X-like, S-like, or X-complex categories in each magnitude bins as well as the corresponding C/X-like:S-like ratio}
         \label{tbl-9}
         \centering
         \begin{tabular}{c c c c c}
            \hline\hline
            Mag bin & C/X-like & S-like & X-complex & C/X-like:S-like ratio \\
             & & & (numbers assigned to C/X-like) & \\
            \hline	
            $<14.00$ & 1 & 11 & 0(0) & 0.09 \\
            14.00-14.99 & 3 & 19 & 5(2) & 0.23 \\
            15.00-15.99 & 1 & 43 & 3(1) & 0.04 \\
            16.00-16.99 & 14 & 65 & 4(2) & 0.24 \\
            17.00-17.99 & 6 & 61 & 10(5) & 0.17 \\
            18.00-18.99 & 12 & 44 & 15(7) & 0.37 \\
            19.00-19.99 & 10 & 39 & 9(4) & 0.32 \\
            20.00-20.99 & 9 & 21 & 7(3) & 0.48 \\
            21.00-21.99 & 1 & 15 & 5(2) & 0.17 \\
            $\geq22.00$ & 1 & 12 & 3(1) & 0.14 \\
            \noalign{\smallskip}
            \hline
         \end{tabular}
\end{table}

\end{document}